\documentclass{article}  
\usepackage{lajolla2006}
\usepackage{graphicx}
\usepackage{epsfig,latexsym}
\usepackage{amssymb}

\newcommand{\be}{\begin{eqnarray}}
\newcommand{\ee}{\end{eqnarray}}

\frompage{000} \topage{000}                                              
\def\rightmark{Temperature-dependence of Quarkonia Correlators}
\def\leftmark{\'A.M\'ocsy }
 
\title{On the Temperature-Dependence of Quarkonia Correlators}
\authors{
{\'Agnes M\'ocsy 
}\\[2.812mm]
{\normalsize
\hspace*{-0.6cm}RIKEN-BNL Research Center, Brookhaven National Laboratory, Upton, NY 11973 
}}
 
\abstract{Here I review the temperature-dependence of heavy quarkonia
  correlators in potential models with three different screened
  potentials, and compare these to the results from lattice QCD. None
  of the potentials investigated yield results consistent with the lattice data, indicating that screening is likely not the mechanism for
  heavy quarkonia suppression. I also discuss a simple toy model, not
  based on temperature-dependent screening, that can reproduce the
  lattice results.}

\PACS{specifications see, e.g.\ {\tt http://www.aip.org/pacs/}}
 
\begin{document}
 
\maketitle
\setcounter{page}{1}

\section{Introduction}

The idea that the melting of heavy quark bound states at the
deconfinement temperature could be considered an unambiguous signal
for deconfinement has led to an intense line of studies.  Originally
it was predicted in \cite{Matsui:1986dk} that color screening in the
deconfined medium would cause the dissolution of the
$J/\psi$. Understanding the modification of the properties of the
different quarkonium states in a hot medium is therefore crucial for
understanding deconfinement. Experiments have been looking for
$J/\psi$ suppression at CERN-SPS and RHIC-BNL
\cite{manuel}. Theoretical studies were mostly phenomenological, and
use potential models as a basic tool. In recent years, first
principle calculations of QCD carried out on the lattice provided new
and unexpected information about quarkonia at high temperatures
\cite{Umeda:2002vr,{Datta:2003ww}}.

\begin{figure}[htbp]
\begin{minipage}[htbp]{5cm}
\epsfig{file=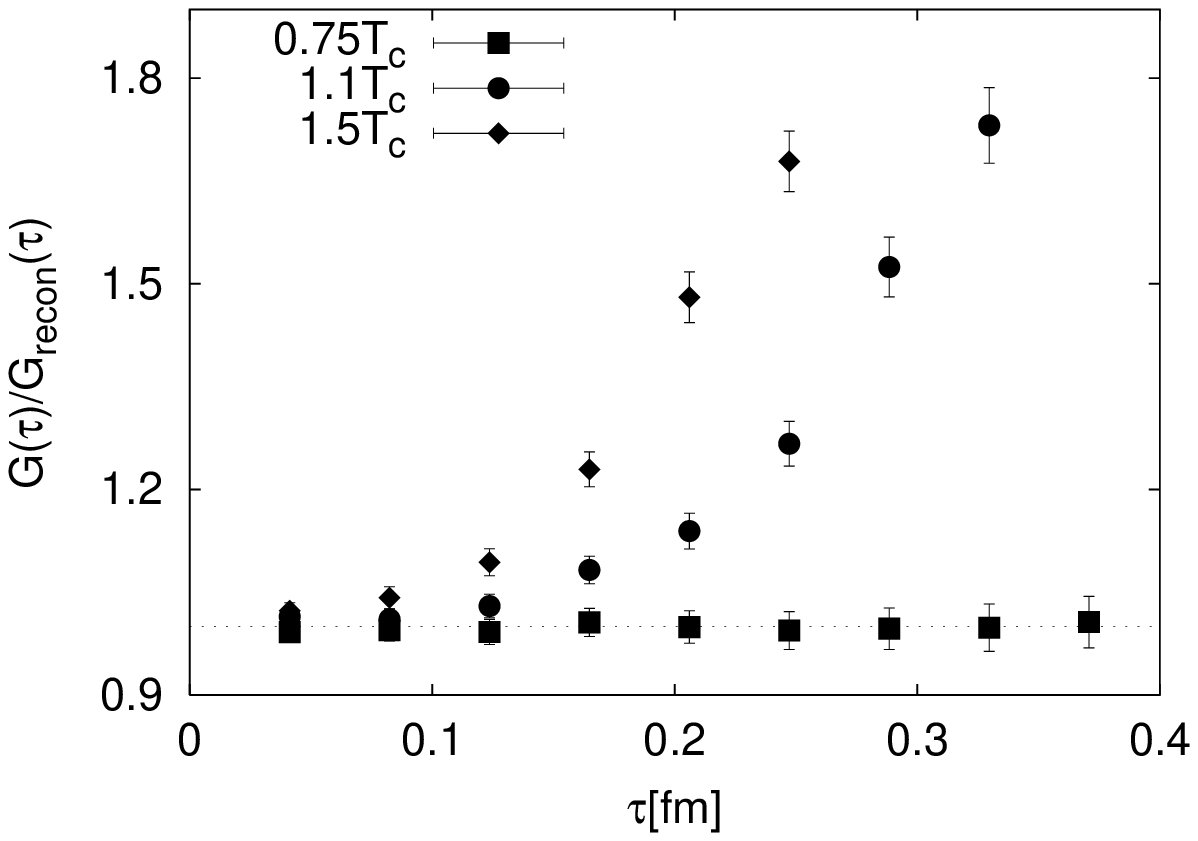,height=46mm}
\end{minipage}
\hspace*{1cm}
\begin{minipage}[htbp]{5cm}
\epsfig{file=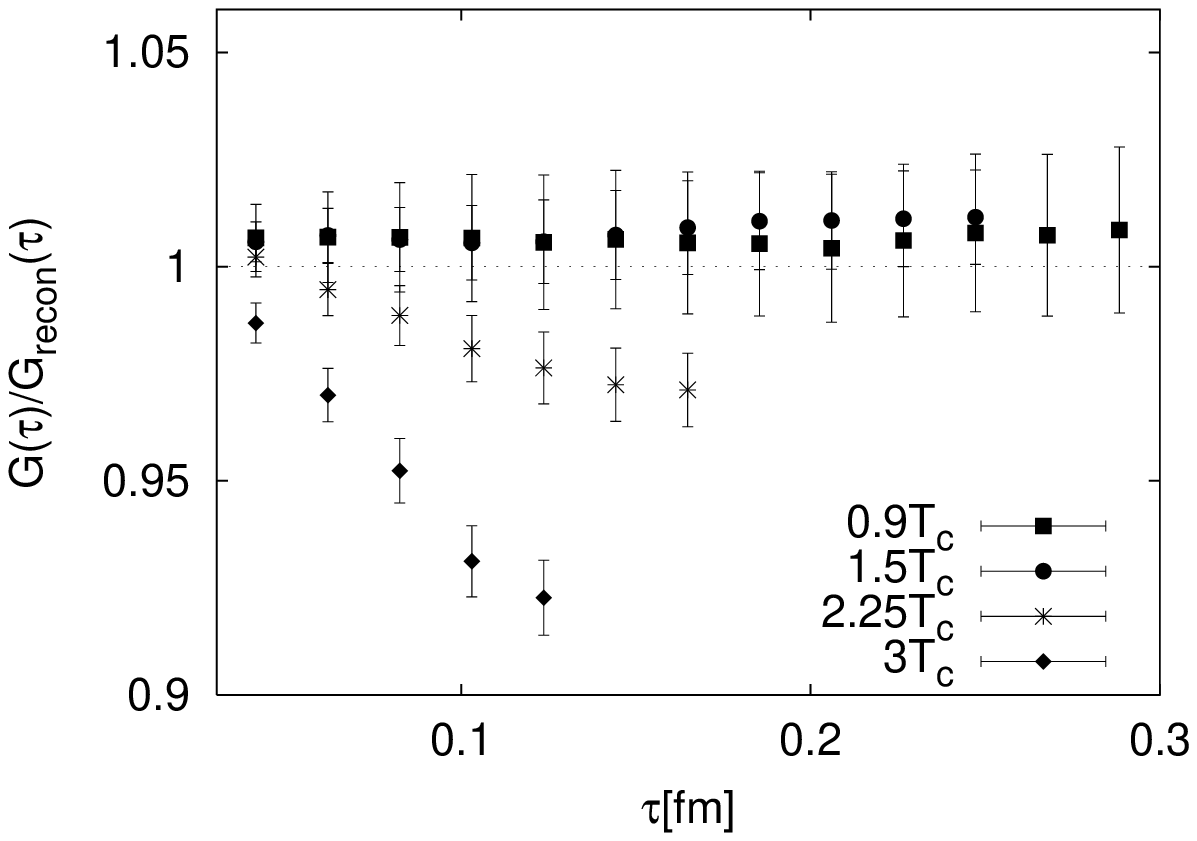,height=46mm}
\end{minipage}
\caption{Temperature dependence of scalar (left panel) and
  pseudo-scalar (right panel) correlators obtained on the lattice (from
  \cite{Datta:2003ww}).}
\label{fig:lattice}
\end{figure}

Correlation functions of hadronic currents $G(\tau,T)$ have been
reliably calculated on the lattice. Any deviation from one of the
ratio \be \frac{G(\tau,T)}{G_{recon}(\tau,T)} = \frac{\int d\omega
  \sigma(\omega,T) K(\tau,\omega,T)}{\int d\omega \sigma(\omega,T=0)
  K(\tau,\omega,T)}
\label{G}
\ee indicates modification of the spectral function $\sigma(\omega,T)$
with temperature. The integration kernel is $K(\tau,\omega,T)
=\cosh{(\omega(\tau-1/2T))}/\sinh{(\omega/2T)}$. Fig.~\ref{fig:lattice}
shows the ratio of correlators (\ref{G}) for the scalar (left panel)
and pseudo-scalar (right panel) charmonium \cite{Datta:2003ww}. In
contradiction with what has been theoretically expected from potential
model calculations (see for instance \cite{Karsch:1987pv}), these
lattice results indicate that the 1S charmonium survives up to 1.5
$T_c$ and the 1P charmonium dissolves by 1.16 $T_c$.  The spectral
functions, extracted from the correlators using the Maximum Entropy
Method, not only reinforce these findings, but also indicate that the
properties of the 1S states do not change up to these temperatures
\cite{Datta:2003ww}.

After the appearance of the lattice data, potential models have been
reconsidered using different temperature dependent potentials
\cite{Shuryak:2003ty,{Wong:2004zr},{Alberico:2005xw},{Mannarelli:2005pa}}.
With these models quarkonium dissociation temperatures in accordance
with the above quoted numbers from the lattice were identified.  In
\cite{Mocsy:2004bv,{Mocsy:2006zd}} however, it has been shown, that
even though potential models with certain screened potentials can
reproduce qualitative features of the lattice spectral function, such
as the survival of the 1S state and the melting of the 1P state, the
temperature dependence of the meson correlators is not
reproduced. Furthermore, the properties of the states determined with
these screened potentials do not seem to reproduce the results
indicated by the lattice spectral functions.

The question is thus whether medium modifications of quarkonia
correlators can be understood via a temperature-dependent
quark-antiquark potential? If yes, what is the potential? And if not,
then what is the relevant mechanism responsible for the dissociation
of quarkonia at high temperatures?
 
Here I review some of the main results of
\cite{Mocsy:2004bv,{Mocsy:2006zd}}, and then present a simple toy
model with no explicit screened potential which provides results that
are consistent with the lattice correlator data. Further developments
are discussed in the Outlook.

\section{Model Spectral Function and Potentials}

In order to make direct comparison with the lattice data we calculate
the ratio of correlators (\ref{G}). We model the finite temperature
spectral function in a given quarkonium channel as the sum of bound
state (resonance) contributions and the perturbative continuum above a
threshold $s_0$, \be \sigma(\omega,T) = \sum_i 2 M_i(T) F_i(T)^2
\delta\left(\omega^2-M_i(T)^2\right) + \frac{3}{8\pi^2}\omega^2
\theta\left(\omega-s_0(T)\right) f \, ,
\label{spft} 
\ee with $f=+1$ and $-1$ in the pseudo-scalar and scalar
channels\footnote{Such a form for the spectral function is justified
  at $T=0$. We assume that it is an appropriate description also at
  finite temperature.}. The mass $M_i$ and the amplitude $F_i$ of the
quarkonium states is determined using potential models.

The essence of potential models is to assume that the interaction
between a heavy quark and antiquark is mediated by a two-body
potential. This assumption is feasible when the quark-antiquark
interaction is instantaneous. The properties of a bound state are
determined by solving the Schr\"odinger equation with this
potential. At zero temperature the Cornell potential seems to have
described quarkonia spectroscopy rather well. At finite temperature,
however, the form of the potential is not known. It is even
questionable whether a temperature-dependent potential is adequate for
the understanding of the properties of quarkonia at finite
temperature.

We calculated the correlators for three different potentials that have
been popular in the literature: First, the screened Cornell potential
\cite{Karsch:1987pv} \be V(r,T)=-\frac{\alpha}{r}e^{-\mu(T)
  r}+\frac{\sigma}{\mu(T)}(1-e^{-\mu(T) r}) \, ,
\label{cornell}
\ee
with parameters described in \cite{Mocsy:2004bv}. 

\begin{figure}[htbp]
\begin{minipage}[htbp]{5cm}
\epsfig{file=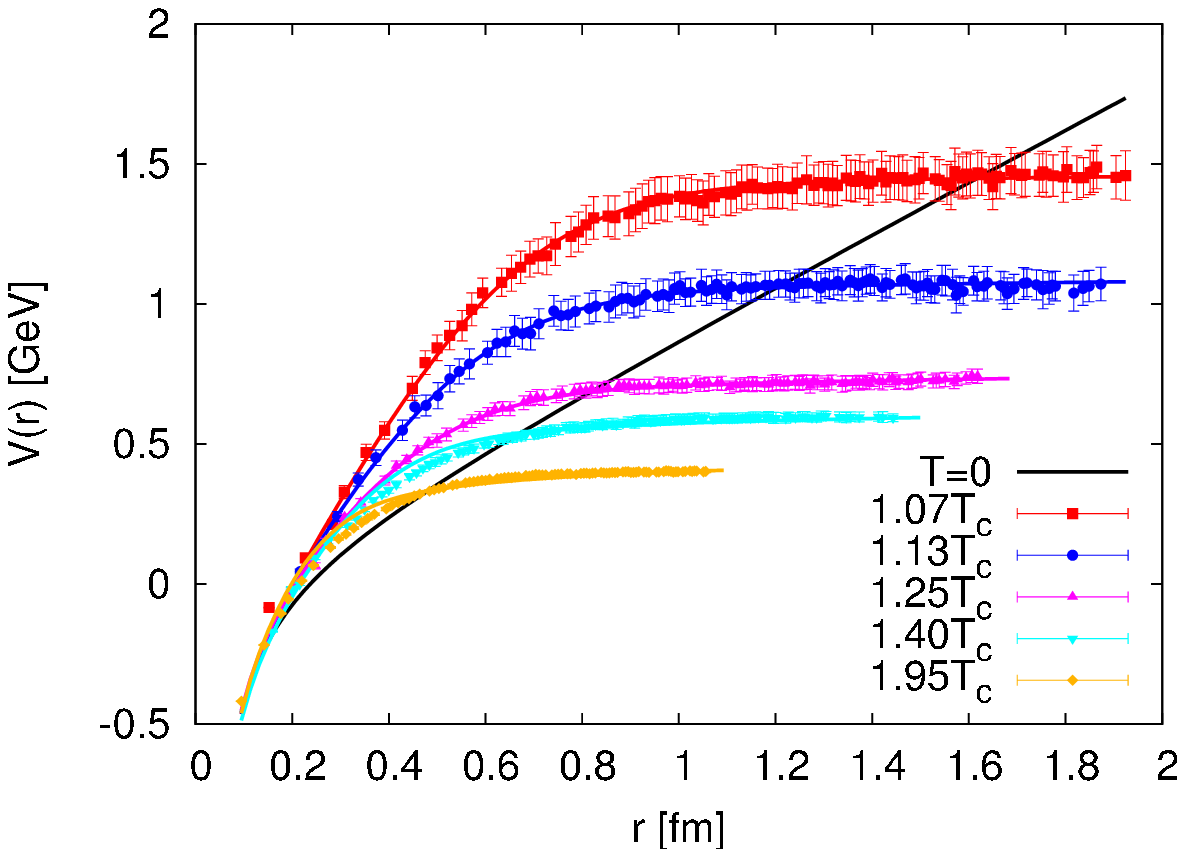,height=40mm}
\end{minipage}
\hspace*{1cm}
\begin{minipage}[htbp]{5cm}
\epsfig{file=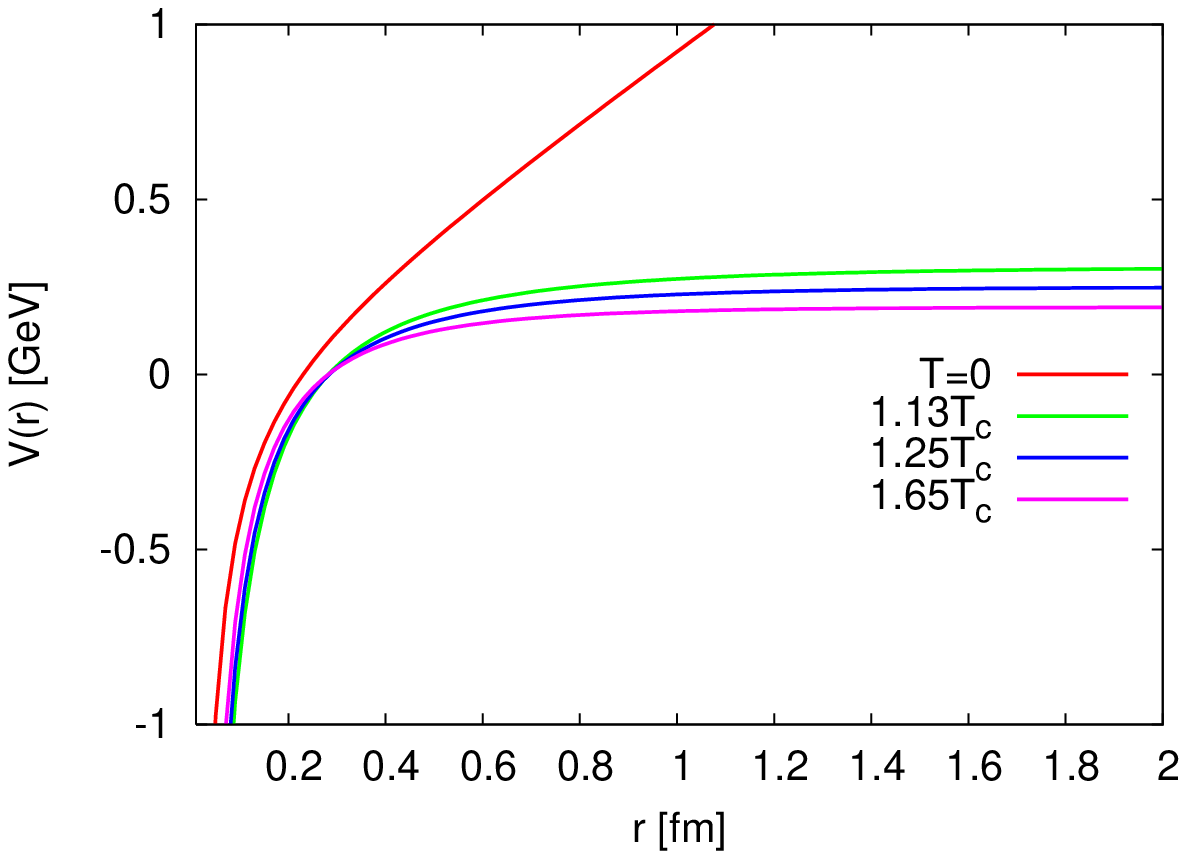,height=43mm}
\end{minipage}
\caption{Lattice internal energy (left panel); Wong-potential  (right panel).} 
\label{fig:pot}
\end{figure}

Second, the internal energy of a heavy quark-antiquark pair as
determined on the lattice \cite{Kaczmarek:2003dp} and identified as
the potential \cite{Shuryak:2003ty}.  Our fit of the internal energy
is shown on the left panel of Fig.~\ref{fig:pot}, and the details of
our parametrization are given in \cite{Mocsy:2004bv}. One should be
aware that in leading order perturbation theory, which is valid at
high temperatures, the potential is equal to the free energy of the
quark-antiquark pair. Beyond leading order there is an entropy
contribution to the free energy and therefore it is conceptually
difficult to identify this with the potential \cite{Petreczky:2005bd}.

Third, we consider a combination of the internal and the free energy
from the lattice that has also been suggested by Wong as potential
\cite{Wong:2004zr}.  This potential is shown on the right panel of
Fig.\ref{fig:pot}. One common feature of all three potentials is that
they incorporate temperature-dependent screening.
%

\section{Results}

\begin{figure}[htbp]
\begin{minipage}[htbp]{5.2cm}
\epsfig{file=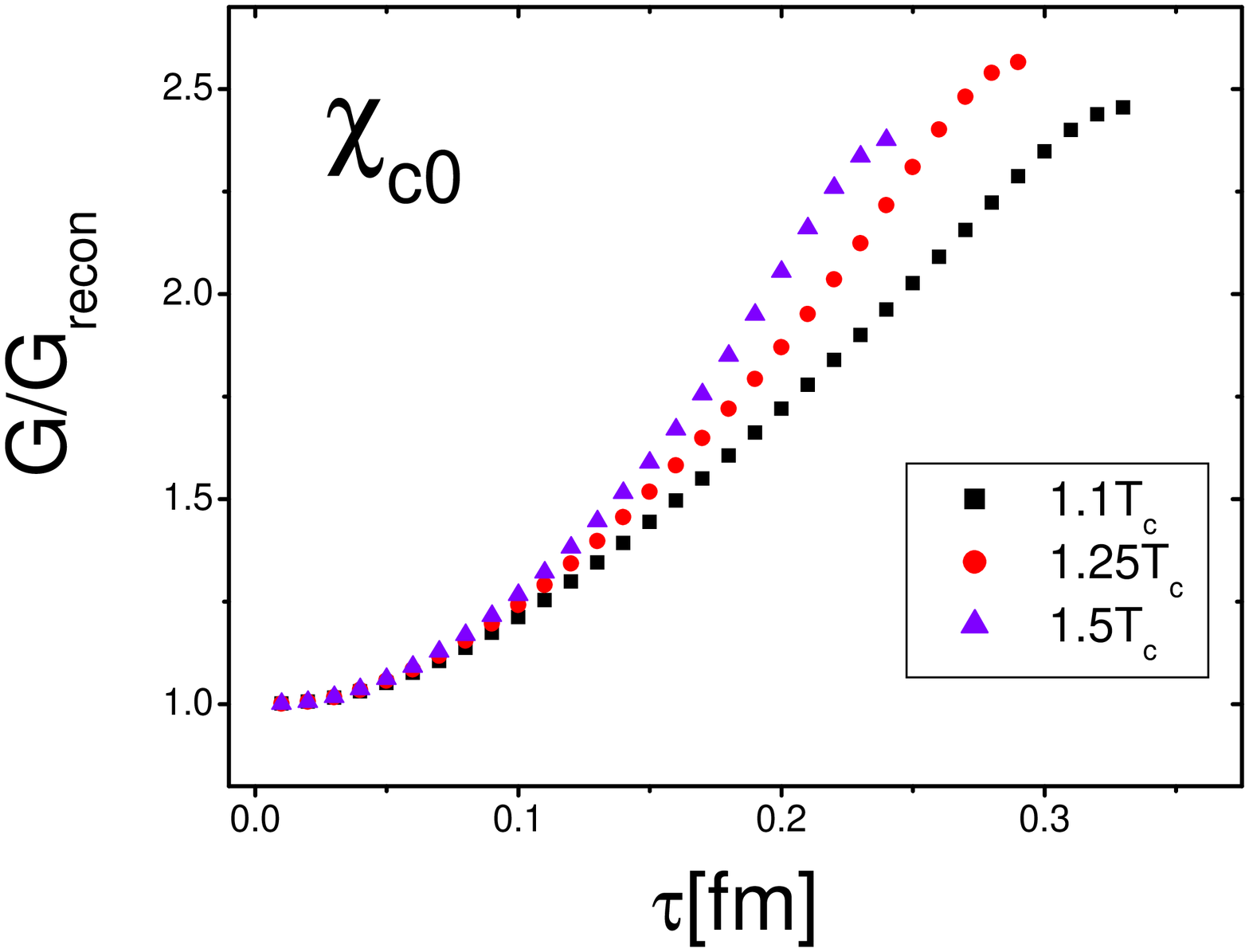,height=46mm}
\end{minipage}
\hspace*{1cm}
\begin{minipage}[htbp]{5cm}
\epsfig{file=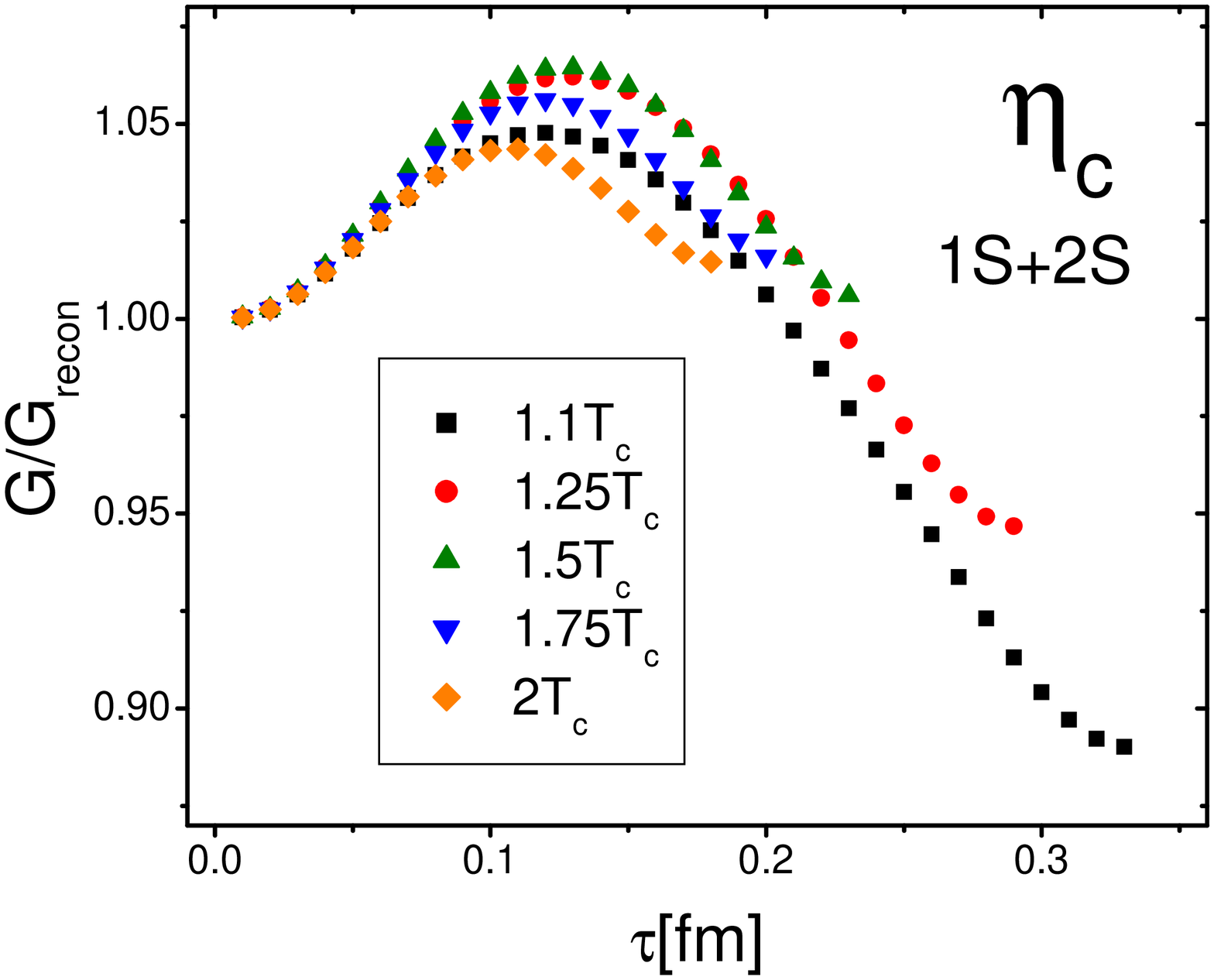,height=46mm}
\end{minipage}
\caption{Temperature dependence of scalar (left panel) and
  pseudo-scalar (right panel) correlators using the screened
  Cornell-potential (\ref{cornell}).}
\label{fig:cornell}
\end{figure}

\begin{figure}[htbp]
\begin{minipage}[htbp]{5cm}
\epsfig{file=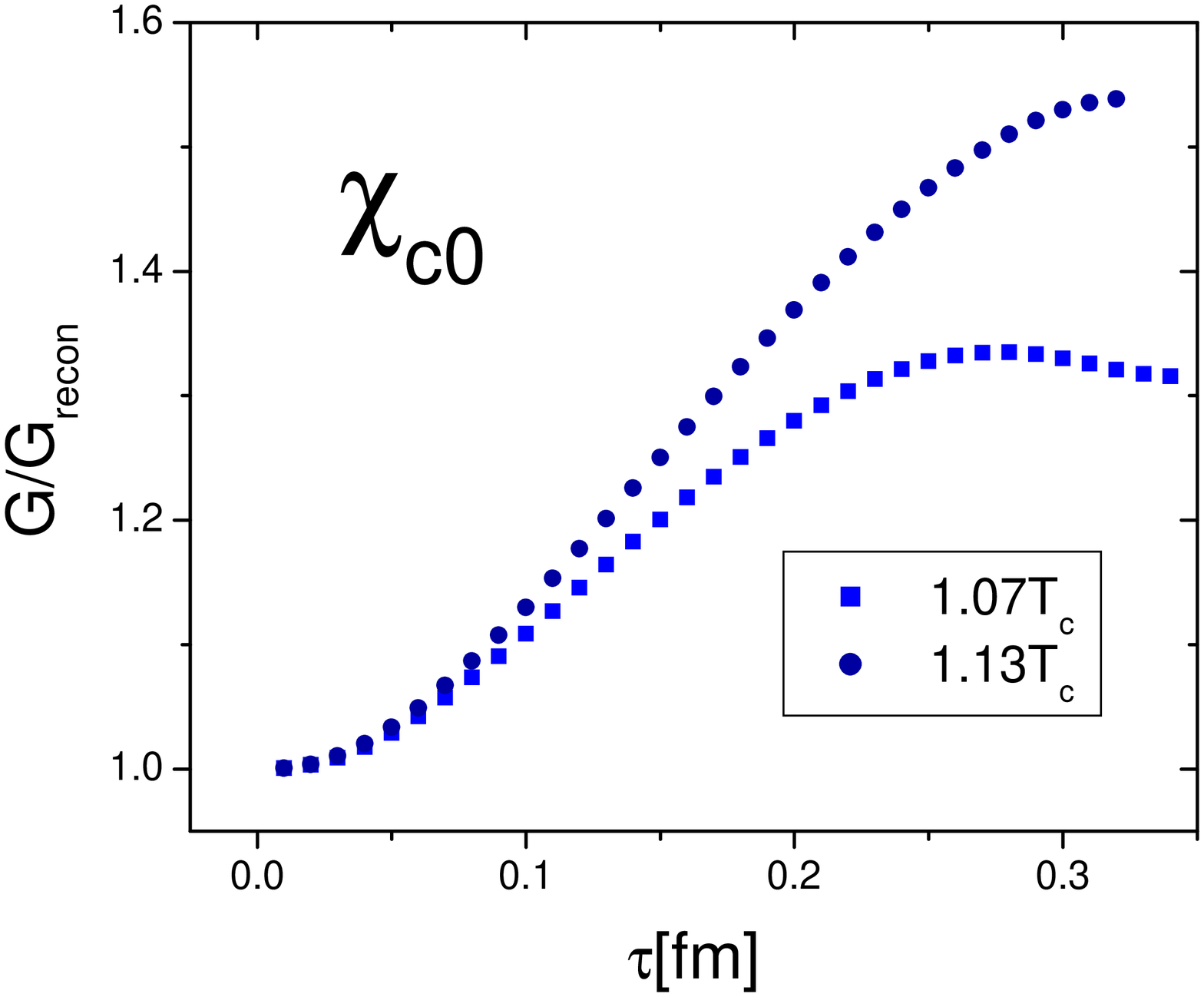,height=46mm}
\end{minipage}
\hspace*{1cm}
\begin{minipage}[htbp]{5cm}
\epsfig{file=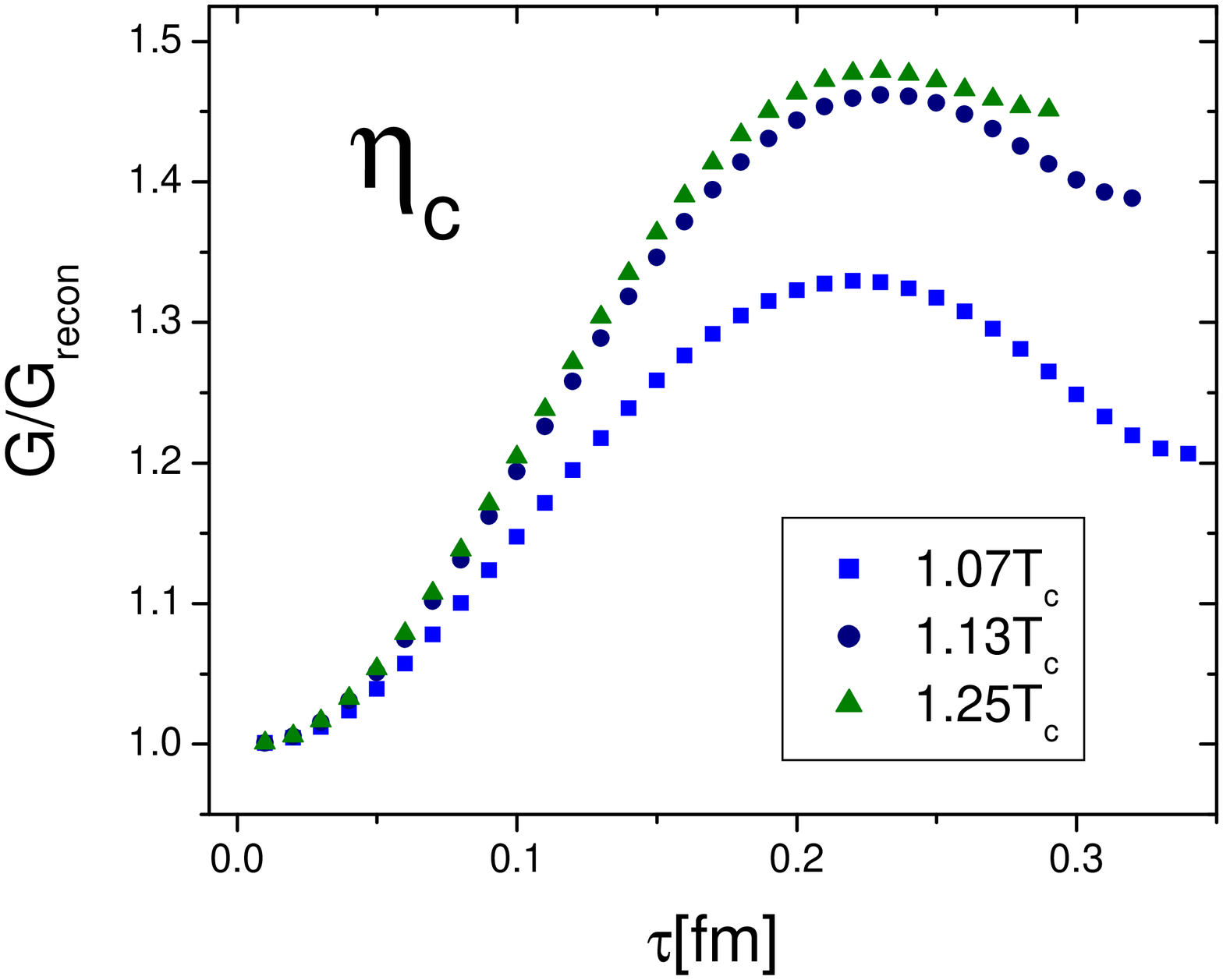,height=46mm}
\end{minipage}
\caption{Temperature dependence of scalar (left panel) and
  pseudo-scalar (right panel) correlators obtained using the lattice
  internal energy as potential.}
\label{fig:internal}
\end{figure}

\begin{figure}[htbp]
\begin{center}
\resizebox{0.55\textwidth}{!}{%
 \includegraphics{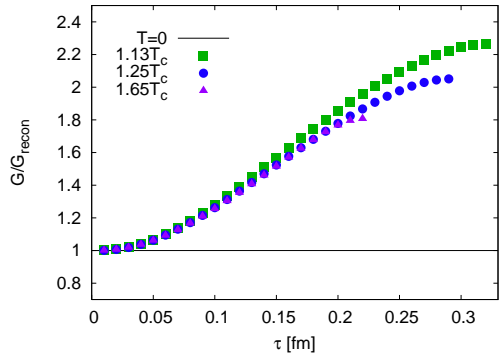}}
\caption{Temperature-dependence of the pseudo-scalar correlator
  obtained using the Wong-potential.}
\label{fig:wong}
\end{center}
\end{figure}

Figs.~\ref{fig:cornell} and \ref{fig:internal} display the ratio of
correlators (\ref{G}) as obtained using the screened Cornell potential
(\ref{cornell}) and the lattice internal energy. The left and right
panels show the results for the scalar $\chi_c$ and the pseudo-scalar
$\eta_c$ for different temperatures.  One can see that the qualitative
behavior for the $\chi_c$ correlator agrees with what is seen on the
lattice (left panel Fig.\ref{fig:lattice}). There is however, no
agreement with the lattice (right panel of Fig.\ref{fig:lattice}) for
the $\eta_c$ correlator. In the model calculations one can identify a
more complex substructure in the $\eta_c$ correlator: The reduction of
the continuum threshold and that the amplitude of the states are
distinguishable contributions (see \cite{Mocsy:2004bv} for details).
The $\eta_c$ correlator obtained using the Wong-potential is shown in
Fig.~\ref{fig:wong}. This also illustrates a large disagreement with
what is seen on the lattice, indicating that the spectral function of
the $\eta_c$ is significantly different than at zero temperature. This further suggests that this state melts near $T_c$
already.  The results for the spectral function presented in
\cite{new} further confirm this statement.

We also analyzed the bottomonium states, and found that in this case
too, the correlators calculated in the potential models cannot
reproduce the lattice results. We refer the interested reader to \cite{Mocsy:2004bv,{new}} .

Clearly, none of these potentials lead to correlators that agree with
the lattice. It is thus a reasonable question to ask whether such
temperature-dependent screened potentials are the right way to
describe modification of quarkonia properties with temperature. As a
first attempt to answer this question consider the following simple
model:

\section{Toy Model} 

Keeping the lattice results in mind, namely that no modification in
the properties of the 1S charmonium compared to the zero temperature
values has been observed up to well above $T_c$, we use for the mass
and decay rate of this state the Particle Data Group values. Also,
since lattice data suggest that higher excited states disappear near
the transition temperature, we "melt" the 2S and 3S states, and also
the 1P state at $T_c$.

\begin{figure}[htbp]
\begin{center}
\resizebox{0.55\textwidth}{!}{%
 \includegraphics{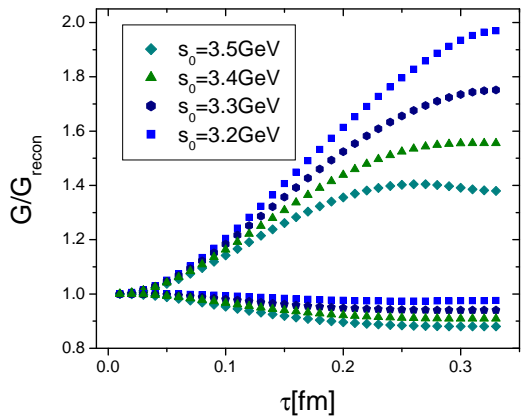}}
\caption{The scalar (upper branch) and pseudo-scalar (lower branch)
  charmonium correlators in the toy model for different values of the
  continuum threshold.}
\label{fig:toy}
\end{center}
\end{figure}

This model does not include temperature dependent screening. The only
parameter is the continuum threshold $s_0$. The main idea is to
compensate for the melting of the higher excited states above $T_c$
with the decrease of the threshold. On Fig. \ref{fig:toy} the
charmonium correlators for the scalar (upper branch) and pseudo-scalar
(lower branch) channels are shown for different values of $s_0$. This
figure illustrates that we can recover the qualitative behavior of the
lattice correlators of Fig.\ref{fig:lattice}: the flatness of the
$\eta_c$ and the increase in the $\chi_c$ correlator.

\section{Outlook} 

We illustrated that potential models utilizing temperature-dependent
screened potentials are not successful in reproducing qualitatively
the lattice results for quarkonium correlators. We further showed that
a simple toy model with no screening is consistent with the
lattice. This model shows that the decrease of the threshold with
increasing temperature can compensate for the melting of the higher
excited states.

To overcome possible errors that could be introduced by our spectral
function Ansatz, we performed a full non-relativistic calculation of
the Green's function \cite{{Mocsy:2006zd},new}, whose imaginary part
provides the quarkonium spectral function. Our results produced for
the different screened potentials again do not show qualitative
agreement with what is seen on the lattice
\cite{{Mocsy:2006zd},new}.

We then conclude that screening is likely not responsible for
quarkonia suppression. This can happen when the time-scale of
screening is not short compared to the time-scale of the heavy quark
motion. Then gluon dissociation becomes the mechanism behind the
dissolution of heavy quarkonia states. This is the topic of our
ongoing investigations.

\section*{Acknowledgments}
This presentation was based on work done in collaboration with P\'eter Petreczky. I thank the
Organizers for a successful workshop.

\vfill\eject

\begin{thebibliography}{9}  

\bibitem{Matsui:1986dk}
T.~Matsui and H.~Satz,
Phys.\ Lett.\ B {\bf 178}, 416 (1986).

\bibitem{manuel} Manuel Calderon , in these proceedings. 

\bibitem{Umeda:2002vr}
  T.~Umeda, K.~Nomura and H.~Matsufuru,
  Eur.\ Phys.\ J.\ C {\bf 39S1}, 9 (2005)
  [arXiv:hep-lat/0211003];
  M.~Asakawa and T.~Hatsuda,
  Phys.\ Rev.\ Lett.\  {\bf 92}, 012001 (2004)
  [arXiv:hep-lat/0308034].

\bibitem{Datta:2003ww}
S.~Datta, F.~Karsch, P.~Petreczky and I.~Wetzorke,
  Nucl.\ Phys.\ Proc.\ Suppl.\  {\bf 119}, 487 (2003)
  [arXiv:hep-lat/0208012]; 
Phys.\ Rev.\ D {\bf 69}, 094507 (2004) [arXiv:hep-lat/0312037].

\bibitem{Karsch:1987pv}
F.~Karsch, M.~T.~Mehr and H.~Satz,
Z.\ Phys.\ C {\bf 37}, 617 (1988).

\bibitem{Shuryak:2003ty}
E.~V.~Shuryak and I.~Zahed,
Phys.\ Rev.\ C {\bf 70}, 021901 (2004) [arXiv:hep-ph/0307267];
Phys.\ Rev.\ D {\bf 70}, 054507 (2004) [arXiv:hep-ph/0403127].

\bibitem{Wong:2004zr}
  C.~Y.~Wong,
  Phys.\ Rev.\ C {\bf 72}, 034906 (2005)
  [arXiv:hep-ph/0408020].

\bibitem{Alberico:2005xw}
  W.~M.~Alberico, A.~Beraudo, A.~De Pace and A.~Molinari,
  Phys.\ Rev.\ D {\bf 72}, 114011 (2005)
  [arXiv:hep-ph/0507084].

\bibitem{Mannarelli:2005pa}
M.~Mannarelli and R.~Rapp,
arXiv:hep-ph/0509310.

\bibitem{Mocsy:2004bv}
  \'A.~M\'ocsy and P.~Petreczky,
  Eur.\ Phys.\ J.\ C {\bf 43}, 77 (2005)
  [arXiv:hep-ph/0411262];
  Phys.\ Rev.\ D {\bf 73}, 074007 (2006)
  [arXiv:hep-ph/0512156].

\bibitem{Mocsy:2006zd}
  \'A.~M\'ocsy and P.~Petreczky,
  arXiv:hep-ph/0606053.

\bibitem{Kaczmarek:2003dp}
O.~Kaczmarek, F.~Karsch, P.~Petreczky and F.~Zantow,
Nucl.\ Phys.\ Proc.\ Suppl.\  {\bf 129}, 560 (2004)
[arXiv:hep-lat/0309121].

\bibitem{Petreczky:2005bd}
  P.~Petreczky,
  Eur.\ Phys.\ J.\ C {\bf 43}, 51 (2005)
  [arXiv:hep-lat/0502008].

\bibitem{new} \'A.~M\'ocsy, P.~Petreczky and J.Casalderrey-Solana, in preparation. 

\end{thebibliography}
\end{document}